%% file: low-complexity-pruned-8-pt-cleaned.tex
\RequirePackage[l2tabu, orthodox]{nag}
\RequirePackage{snapshot}

\documentclass[10pt,onecolumn]{extarticle}

\makeatletter
\if@twocolumn
  \usepackage[dvips,letterpaper,top=0.5in, bottom=0.5in, left=0.75in, right=0.5in,includefoot,heightrounded]{geometry}
\else
  \usepackage[dvips,letterpaper,margin=1in,includefoot,heightrounded]{geometry}
\fi

\usepackage{srcltx}

\usepackage{bm}

\usepackage{cite}
\usepackage{url}
\usepackage[cmex10]{amsmath}
\usepackage{amssymb,amsfonts}
\usepackage{epstopdf}
\usepackage{array}
\usepackage{epsfig}
\usepackage[usenames,dvipsnames]{color}
\usepackage{flushend}

\usepackage{graphics}
\usepackage{graphicx}
\usepackage[tight,footnotesize]{subfigure}
\usepackage{stfloats}
\usepackage{flushend}
\usepackage{setspace}

\usepackage{fullpage}
\usepackage[normalem]{ulem}

\usepackage{multirow}

\usepackage{array}

\usepackage{booktabs}

\usepackage{mathptmx}

\usepackage{color}

\usepackage{hyperref}

\usepackage[sc,tiny,center]{titlesec}

\usepackage{microtype}



\newcommand{\printtitle}{%
\makeatletter
\if@twocolumn

\twocolumn[%
  \maketitle
  \begin{onecolabstract}
    \myabstract
  \end{onecolabstract}
  \begin{center}
    \small
    \textbf{Keywords}
    \\\medskip
    \mykeywords
  \end{center}
  \bigskip
]
\saythanks
\else
  \maketitle
  \begin{abstract}
    \myabstract
  \end{abstract}
  \begin{center}
    \small
    \textbf{Keywords}
    \\\medskip
    \mykeywords
  \end{center}
  \bigskip
  \onehalfspacing
\fi
}

\title{%
Low-complexity Pruned 8-point DCT Approximations for Image Encoding
}

\author{%
V\'itor~A.~Coutinho
\thanks{
V\'itor~A.~Coutinho
is with
the
Signal Processing Group,
Dept. of Statistics,
Federal University of Pernambuco~(UFPE)
and
the graduate program in Electrical Engineering~(PPGEE),
UFPE, Recife, PE, Brazil
(e-mail: vitor.andrade.coutinho@gmail.com).
}
\and
Renato~J.~Cintra\thanks{
R.~J.~Cintra
is with the
Signal Processing Group,
Departamento de Estat\'istica,
Universidade Federal de Pernambuco,
Recife, PE, Brazil;
and
the
Department of Electrical and Computer Engineering,
University of Calgary, Calgary, AB, Canada
(e-mail: rjdsc@de.ufpe.br).
}
\and
F\'abio~M.~Bayer%
\thanks{%
F\'abio~M.~Bayer
is with the
Departamento de Estat\'istica
and Laborat\'orio de Ci\^encias Espaciais de Santa Maria (LACESM),
Universidade Federal de Santa Maria,
Santa Maria, RS, Brazil
(e-mail: bayer@ufsm.br).}
\and
Sunera~Kulasekera
\and
Arjuna~Madanayake%
\thanks{%
Sunera Kulasekera
and
Arjuna Madanayake
are with the
Department of Electrical and Computer Engineering,
The University of Akron, Akron, OH, USA
(e-mail: arjuna@uakron.edu)}.
}

\date{}

\newcommand{\myabstract}{%
Two multiplierless pruned 8-point
discrete cosine transform (DCT)
approximation are presented.
Both transforms present lower arithmetic complexity than state-of-the-art methods.
The performance of such new methods was assessed
in the image compression context.
A JPEG-like simulation was performed,
demonstrating the
adequateness and competitiveness
of the introduced methods.
Digital VLSI implementation in CMOS technology
was also considered.
Both presented methods were realized in Berkeley Emulation Engine (BEE3).
}

\newcommand{\mykeywords}{%
Approximate discrete cosine transform,
pruning,
pruned DCT,
HEVC
}

\begin{document}

\printtitle

\section{Introduction}

Orthogonal transforms are useful tools in many scientific applications~\cite{ahmed1975}.
In particular,
the discrete cosine transform (DCT)~\cite{rao1990discrete} plays an important role in digital signal processing. Due to the energy compaction property, which are close related to the Karuhnen-Lo\`eve
transform (KLT)~\cite{ahmed1974}, the DCT is often used in data compression~\cite{britanak2007discrete}. In fact, the DCT is adopted in many image and video compression standards, such as
JPEG~\cite{Wallace1992},
MPEG~\cite{Gall1992, roma2007hybrid, mpeg2},
H.261~\cite{h261, Liou1990},
H.263~\cite{h263, roma2007hybrid},
H.264/AVC~\cite{wiegand2003, h264},
and the high efficiency video coding (HEVC)~\cite{hevc,hevc1}. Due to its range of applications, efforts have been made over decades to develop fast algorithms to compute the DCT efficiently~\cite{britanak2007discrete}.

The theoretical multiplicative complexity minimum for the $N$-point DCT was derived by Winograd in~\cite{winograd1980}. For the \mbox{8-point} DCT, such minimum consists of 11 multiplications, which is achieved by Loeffler DCT algorithm~\cite{Loeffler1989}.
Consequently,
new algorithms that could
further and significantly
reduce the computational cost of
the exact DCT computation are not trivially attainable.
In this scenario,
low-complexity DCT approximations have been proposed for image and video compression applications,
such as
the classical signed DCT (SDCT)~\cite{haweel2001},
the Lengwehasatit-Ortega DCT approximation (LODCT)~\cite{lengwehasatit2004scalable},
the Bouguezel-Ahmad-Swamy (BAS) series~\cite{bas2008, bas2009, bas2013}, and
algorithms based on integer functions~\cite{cb2011, bc2012, Cintra2014-sigpro}.
Such methods are often multiplierless operations leading to hardware realizations that offer adequate
trade-offs between accuracy and complexity~\cite{Potluri2013}.

In some applications, most of the useful signal information is concentrated in the lower DCT coefficients.
Furthermore, in data compression applications,
high frequency coefficients are often zeroed by
the quantization process~\cite[p.~586]{Malepati2010}.
Then,
computational
savings may be attained
by not computing DCT higher coefficients.
This approach is called \emph{pruning} and was first applied for the discrete Fourier transform (DFT) by Markel~\cite{Markel1971}.
In this case, pruning consists of discarding input vector coefficients---time-domain pruning---and operations involving them are not computed. Alternatively, one can discard output vector coefficients---frequency-domain pruning.
The well-known Goertzel algorithm for single component DFT
computation~\cite{Oppenheim2010,kim2011islanding,carugati2012variable}
can be understood in this latter sense. Frequency-domain pruning has been recently considered for mixed-radix FFT algorithms~\cite{wang2012generic}, cognitive radio design~\cite{airoldi2010energy}, and wireless communications~\cite{whatmough2012vlsi}.

The DCT pruning algorithm was proposed by Wang~\cite{wang1991pruning}.
Due to the DCT energy compaction property,
pruning has been applied in frequency-domain by discarding transform-domain coefficients with
the least amounts of energy~\cite{wang1991pruning}.
In the context of wireless image sensor networks,
Lecuire \emph{et~al.} extended the method for the \mbox{2-D} case~\cite{Lecuire2012}. A pruned DCT approximation was proposed in~\cite{kouadria2013low} based on the DCT approximation method presented in~\cite{bas2013}.

In~\cite{meher2014efficient},
an alternative architecture for HEVC was proposed.
Such method maintains the wordlength fixed
by means of discarding least significant bits, minimizing the computation
at the expense of wordlength truncation.
Although conceptually different, the approach described in~\cite{meher2014efficient} was also referred to as `pruning', in contrast with the classic and more common usage of this terminology. In the current work, we employ `pruning' in the same sense as in~\cite{wang1991pruning}.

The aim of this work is to present new pruned DCT approximations with lower arithmetic complexity
adequate for image coding and compression.
A comprehensive
arithmetic complexity assessment
among several methods is presented.
A JPEG-like image
compression simulation based on the proposed tools
is applied to several images for a performance evaluation.
VLSI architectures for the new presented methods are also introduced.

\section{Mathematical Background}
\subsection{Discrete Cosine Transform}

There are eight types of DCT~\cite{britanak2007discrete}. The most popular one is the type II DCT or DCT-II, which is the best approximation for the KLT for highly correlated signals, being widely employed for data  compression~\cite{Rao2001}.
In this work we refer to the DCT-II simply as DCT.

Let
$\mathbf{x} = \left[x_0~x_1~\ldots~x_{N-1}\right]^\top$
be an input $N$-point vector.
The DCT of $\mathbf{x}$
is
defined as the output vector
$\mathbf{X} = \left[X_0~X_1~\ldots~X_{N-1}\right]^\top$,
whose components are given by
\begin{align*}
X_k \!= \!
\alpha_k
\sqrt{\frac{2}{N}}
\sum_{n=0}^{N-1}
x_n
\cos
\left[
\frac{\left(n+\frac{1}{2}\right)k\pi}{N}
\right]
\!,
k = 0,1,\ldots, N\!-\!1
,
\end{align*}
where
\begin{equation*}
\alpha_k=
\begin{cases}
	\frac{1}{\sqrt{2}}, & \text{if}~k=0, \\
	1, & \text{otherwise.}
\end{cases}
\label{alpha}
\end{equation*}
Alternatively,
above expression can be expressed
in matrix format according to:
\begin{equation}
\label{dct_matrix}
\mathbf{X} = \mathbf{C}\cdot \mathbf{x}
,
\end{equation}
where
$\mathbf{C}$ is the $N \times N$ DCT matrix,
whose entries are given by
\begin{align*}
c_{k,n}=\alpha_k\cdot\sqrt{2/N}\cdot \cos\left[(n+1/2)k\pi/N\right], \\
k,n=0,1,\ldots,N-1.
\end{align*}
Matrix $\mathbf{C}$ is orthogonal, i.e.,
it satisfies $\mathbf{C}^{-1}=\mathbf{C}^\top$.
Thus,
the inverse transformation is $\mathbf{x} = \mathbf{C}^\top\cdot \mathbf{X}$.
Given an $N\times N$ matrix $\mathbf{A}$,
its forward \mbox{2-D} DCT is defined as the transform-domain matrix $\mathbf{B}$ furnished by
\begin{equation*}
\mathbf{B} = \mathbf{C}\cdot \mathbf{A}\cdot \mathbf{C}^\top.
\end{equation*}
By using orthogonality property, the reverse \mbox{2-D} $N$-point DCT is given by
\begin{equation*}
\mathbf{A} = \mathbf{C}^\top\cdot \mathbf{B}\cdot \mathbf{C}.
\end{equation*}
The forward \mbox{2-D} DCT can be computed by
eight column-wise calls
of the \mbox{1-D} DCT to $\mathbf{A}$;
then the resulting intermediate matrix
is submitted to eight row-wise calls of the \mbox{1-D} DCT.

\subsection{DCT Approximations}

A DCT approximation~$\hat{\mathbf{C}}$
is a matrix with similar properties to the DCT,
generally requiring lower computational cost.
In general,
an approximation
is constituted by
the product
$\hat{\mathbf{C}}=\mathbf{S}\cdot \mathbf{T}$,
where
$\mathbf{T}$ is a low-complexity matrix
and
$\mathbf{S}$ is a scaling diagonal matrix,
which effects orthogonality or quasi-orthogonality~\cite{Cintra2014-sigpro}.
In the image compression context,
matrix $\mathbf{S}$ does not contribute
with any extra computation,
since it can be merged
into
the quantization step of
usual compression algorithms~\cite{bas2008,bas2009,bc2012,Potluri2013}.

Current literature contains several good approximations.
In this work,
we separate the following approximations for analysis:
(i)~the SDCT~\cite{haweel2001},
which is a classic DCT approximation;
(ii)~the LODCT~\cite{lengwehasatit2004scalable},
due to its comparatively high performance;
(iii)~the BAS series of approximations~\cite{bas2008,bas2009,bas2013};
(iv)~the rounded DCT (RDCT)~\cite{cb2011};
(v)~the modified RDCT (MRDCT)~\cite{bc2012}, which has the lowest arithmetic complexity in the literature;
and
(vi)~transformation matrices
$\mathbf{T}_4$, $\mathbf{T}_5$ and $\mathbf{T}_6$
introduced in~\cite{Cintra2014-sigpro},
which are fairly recently proposed methods
with good performance and low complexity.

\subsection{Pruning Exact and Approximate DCT}

DCT pruning consists in
discarding selected input or output vector components in~\eqref{dct_matrix},
thus avoiding computations that require them.
Mathematically,
it corresponds to
the elimination
of rows or columns
of $\mathbf{C}$.
Such operations results in a possibly rectangular submatrix
of the original matrix $\mathbf{C}$.
Pruning is often realized in frequency-domain
by means of computing only the $K<N$ transform coefficients
that retain more energy.
For the DCT,
this corresponds
to the
low index coefficients
of the \mbox{1-D} transform
and the upper-left coefficients of \mbox{2-D}.
The new transformation matrix for this particular pruning method is the $K\times N$ matrix $\mathbf{C}_{\langle K\rangle}$
given by
\begin{align}
\mathbf{C}_{\left<K\right>} =
\begin{bmatrix}
c_{0,0}&c_{0,1}&\cdots&c_{0,N-1} \\
c_{1,0}&c_{1,1}&\cdots&c_{1,N-1} \\
\vdots&\vdots&\ddots& \vdots \\
c_{K-1,0}&c_{K-1,1}&\cdots&c_{K-1,N-1} \\
\end{bmatrix}
\label{c_k}
\end{align}
Therefore,
the
\mbox{2-D} pruned DCT is computed as follows:
\begin{equation}
\tilde{\mathbf{B}}=\mathbf{C}_{\langle K\rangle}\cdot \mathbf{A}\cdot \mathbf{C}_{\langle K\rangle}^\top.
\label{pruned_2d}
\end{equation}
The resulting matrix $\tilde{\mathbf{B}}$
is a square matrix of size $K \times K$.
Lecuire \emph{et~al.}~\cite{Lecuire2012}
showed that the
$K \times K$ square pattern at the upper-right corner
leads to
a better energy-distortion trade-off
when compared to the alternative triangle pattern~\cite{Makkaoui2010}.

Pruning can be applied to DCT approximations
by means of discarding matrix rows of $\mathbf{T}$
that correspond to low-energy coefficients.
Thus, the pruned transformation is expressed according to:
\begin{align}
\label{t_k}
\mathbf{T}_{\langle K\rangle}
=
\begin{bmatrix}
t_{0,0}&t_{0,1}&\cdots&t_{0,N-1} \\
t_{1,0}&t_{1,1}&\cdots&t_{1,N-1} \\
\vdots&\vdots&\ddots& \vdots \\
t_{K-1,0}&t_{K-1,1}&\cdots&t_{K-1,N-1} \\
\end{bmatrix}
.
\end{align}
Above transformation furnishes the pruned approximation DCT:
$\hat{\mathbf{C}}_{\langle K\rangle}=\mathbf{S}_{\langle K\rangle}\cdot \mathbf{T}_{\langle K\rangle}$,
where
$\mathbf{S}_{\langle K\rangle} = \sqrt{\operatorname{diag}\left[\mathbf{T}_{\langle K\rangle}\cdot \mathbf{T}_{\langle K\rangle}^\top\right]^{-1}}$
and $\operatorname{diag}(\cdot)$ returns a diagonal matrix with the diagonal elements of its argument.

\section{Proposed Pruned DCT Approximations}
\label{proposed_pruned_dct}

\subsection{Pruned LODCT}

\begin{figure}
\centering
\input{new_ortegak4.pstex_t}\label{fig:ortegak4}
\caption{Fast algorithm for $\mathbf{W}_{\langle 4\rangle}$.}
\label{fig:ortega_k4}
\end{figure}
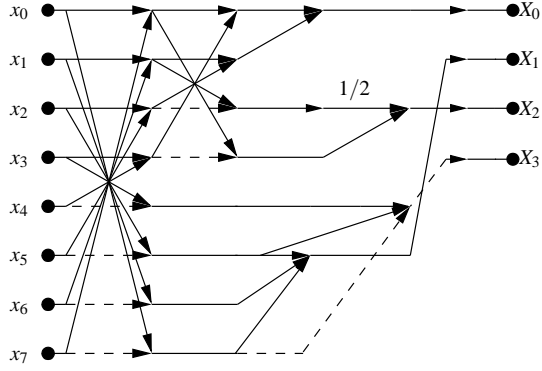

Presenting good performance at image compression and energy compaction,
the LODCT~\cite{lengwehasatit2004scalable}
is associated to the
following transformation matrix:
\begin{align}
\label{ortega}
\mathbf{W}=
\begin{bmatrix}
1 & 1 & 1 & 1 & 1 & 1 & 1 & 1 \\
1 & 1 & 1 & 0 & 0 &-1 &-1 &-1 \\
1 & \frac{1}{2}  & -\frac{1}{2} & -1 &-1 & -\frac{1}{2} & \frac{1}{2} & 1 \\
1 & 0 & -1 & -1 & 1 & 1 & 0 & -1 \\
1 &-1 &-1 & 1 & 1 &-1 &-1 & 1 \\
1 &-1 & 0 & 1 & -1 & 0 & 1 & -1 \\
\frac{1}{2} & -1 & 1 & -\frac{1}{2} & -\frac{1}{2} & 1 & -1 & \frac{1}{2} \\
0 & -1 & 1 & -1 & 1 & -1 & 1 & 0
\end{bmatrix}
.
\end{align}
The implied approximation
matrix is furnished by
$\hat{\mathbf{C}}=\mathbf{S}\cdot \mathbf{W}$,
where
$\mathbf{S}=\operatorname{diag}\left(\frac{1}{\sqrt{8}} ,\frac{1}{\sqrt{6}} ,\frac{1}{\sqrt{5}} ,\frac{1}{\sqrt{6}} ,\frac{1}{\sqrt{8}} ,\frac{1}{\sqrt{6}}, \frac{1}{\sqrt{5}},\frac{1}{\sqrt{6}} \right)$.
The fast algorithm for $\mathbf{W}$
requires 24~additions and 2~bit-shift operations~\cite{lengwehasatit2004scalable}.
By analyzing fifty $512 \times 512$ standard images from a public bank~\cite{USC_database},
we noticed that the
LODCT can concentrate
$\approx\!98.98\%$
of the average total image energy
in the transform-domain upper-left square of size $K=4$.
Thus,
computational savings
can be achieved by computing only the 16~lower frequency coefficients.
Above facts lead to the following pruned transformation based on the LODCT:
\begin{align}
\label{ortega_pruned}
\mathbf{W}_{\langle 4\rangle}=
\begin{bmatrix}
1 & 1 & 1 & 1 & 1 & 1 & 1 & 1 \\
1 & 1 & 1 & 0 & 0 &-1 &-1 &-1 \\
1 & \frac{1}{2}  & -\frac{1}{2} & -1 &-1 & -\frac{1}{2} & \frac{1}{2} & 1 \\
1 & 0 & -1 & -1 & 1 & 1 & 0 & -1
\end{bmatrix}
.
\end{align}
The corresponding pruned approximation is:
$\hat{\mathbf{C}}_{\langle 4\rangle} = \mathbf{S}_{\langle 4\rangle}\cdot \mathbf{W}_{\langle 4\rangle}$,
where
$\mathbf{S}_{\langle 4\rangle}=\operatorname{diag}\left(\frac{1}{\sqrt{8}} ,\frac{1}{\sqrt{2}} ,\frac{1}{2} ,\frac{1}{\sqrt{2}}\right)$.
A fast algorithm flow graph for the pruned \mbox{1-D} LODCT is shown in
\figurename~\ref{fig:ortega_k4},
requiring only 18~additions and 1~bit-shift operation.

\subsection{Pruned MRDCT}

The MRDCT presents
the lowest arithmetic complexity
among
the meaningful DCT approximation in literature~\cite{bc2012}.
It is associated to the following low-complexity matrix:
\begin{align*}
\label{mrdct}
\mathbf{M}=
\begin{bmatrix}
1 & 1 & 1 & 1 & 1 & 1 & 1 & 1 \\
1 & 0 & 0 & 0 & 0 & 0 & 0 &-1 \\
1 & 0 & 0 &-1 &-1 & 0 & 0 & 1 \\
0 & 0 &-1 & 0 & 0 & 1 & 0 & 0 \\
1 &-1 &-1 & 1 & 1 &-1 &-1 & 1 \\
0 &-1 & 0 & 0 & 0 & 0 & 1 & 0 \\
0 &-1 & 1 & 0 & 0 & 1 &-1 & 0 \\
0 & 0 & 0 &-1 & 1 & 0 & 0 & 0
\end{bmatrix}
.
\end{align*}
This matrix
furnishes the DCT approximation given by
$\tilde{\mathbf{C}}= \mathbf{D}\cdot \mathbf{M}$,
where
$\mathbf{D}=\operatorname{diag}\left(\frac{1}{\sqrt{8}} ,\frac{1}{\sqrt{2}} ,\frac{1}{2} ,\frac{1}{\sqrt{2}} ,\frac{1}{\sqrt{8}} ,\frac{1}{\sqrt{2}}, \frac{1}{2},\frac{1}{\sqrt{2}} \right)$.
Its fast algorithm presents only 14~additions~\cite{bc2012}.

\begin{figure}
\centering
\input{mrdct_k6.pstex_t}
\caption{Fast algorithm for $\mathbf{M}_{\langle 6\rangle}$.}
\label{fig:mrdct_k6}
\end{figure}
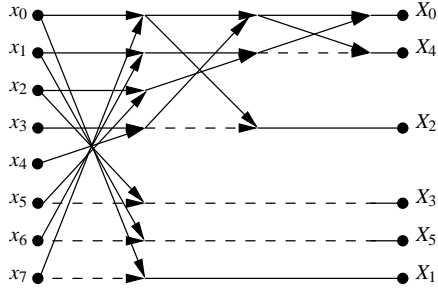

We submitted the same above-mentioned
set of fifty images to the MRDCT and
we noticed that
it can concentrate $\approx\!99.34\%$ of the total average energy in the upper-left square of size $K=6$.
Then, we propose the following pruned $6\times 8$ matrix:
\begin{align*}
\mathbf{M}_{\langle 6\rangle}=
\begin{bmatrix}
1 & 1 & 1 & 1 & 1 & 1 & 1 & 1 \\
1 & 0 & 0 & 0 & 0 & 0 & 0 &-1 \\
1 & 0 & 0 &-1 &-1 & 0 & 0 & 1 \\
0 & 0 &-1 & 0 & 0 & 1 & 0 & 0 \\
1 &-1 &-1 & 1 & 1 &-1 &-1 & 1 \\
0 &-1 & 0 & 0 & 0 & 0 & 1 & 0
\end{bmatrix}
.
\end{align*}
The DCT approximation is given by
$\tilde{\mathbf{C}}_{\langle 6\rangle}= \mathbf{D}_{\langle 6\rangle}\cdot\mathbf{M}_{\langle 6\rangle}$,
where
$\mathbf{D}_{\langle 6\rangle}=\operatorname{diag}\left(\frac{1}{\sqrt{8}} ,\frac{1}{\sqrt{2}} ,\frac{1}{2} ,\frac{1}{\sqrt{2}} ,\frac{1}{\sqrt{8}} ,\frac{1}{\sqrt{2}} \right)$.
A fast algorithm for the pruned \mbox{1-D} MRDCT is shown in \figurename~\ref{fig:mrdct_k6},
requiring
only 12~additions.

\subsection{Inverse Transformation}

The direct 2-D transformation for the pruned LODCT furnishes a $4 \times 4$ matrix $\mathbf{B} = \hat{\mathbf{C}}_{\langle 4 \rangle} \cdot \mathbf{A} \cdot \hat{\mathbf{C}}_{\langle 4 \rangle}^\top$. These 16 coefficients retain most of the signal energy as represented by the original LODCT. Then, the inverse 2-D LODCT can be invoked as shown below:
\begin{equation*}
\hat{\mathbf{A}}
=
\hat{\mathbf{C}}^\top \cdot
\begin{bmatrix}
\mathbf{B} & \mathbf{0} \\
\mathbf{0} & \mathbf{0}
\end{bmatrix}_{8 \times 8}
\cdot \hat{\mathbf{C}}
,
\end{equation*}
where $\hat{\mathbf{C}}$ is the non-pruned LODCT matrix.
Therefore, matrix $\hat{\mathbf{A}}$ is an approximation of $\mathbf{A}$.
However, taking into account the zero element locations and the fact the $\hat{\mathbf{C}}_{\langle 4 \rangle}$ consists of the first four rows of $\hat{\mathbf{C}}$, we have the following alternative expression: $\hat{\mathbf{A}}=\hat{\mathbf{C}}_{\langle 4 \rangle}^\top \cdot \mathbf{B} \cdot \hat{\mathbf{C}}_{\langle 4 \rangle}$,
where $\hat{\mathbf{C}}_{\langle 4 \rangle}$ is the proposed  $4 \times 8$ pruned LODCT matrix.
Furthermore, matrix $\hat{\mathbf{C}}_{\langle 4 \rangle}^\top$ is the Moore-Penrose generalized inverse of
$\hat{\mathbf{C}}_{\langle 4 \rangle}$~\cite[p. 363]{Bernstein2009} .
The computation of the inverse pruned MRDCT follows the same rationale as described above.

\subsection{Complexity Assessment}

On account of~\eqref{pruned_2d},
we have that
\mbox{2-D} pruned approximate DCT
is computed after eight column-wise calls of
the \mbox{1-D} pruned approximate DCT
and $K$~row-wise calls of the same transformation.
Let
$\operatorname{A}_\text{1-D}\left(\mathbf{T}_{\langle K\rangle}\right)$
be
the additive complexity of $\mathbf{T}_{\langle K\rangle}$.
Thus,
the additive complexity of the
\mbox{2-D} pruned approximate DCT is given by:
\begin{equation}
\begin{split}
\operatorname{A}_\text{2-D}\left(\mathbf{T}_{\langle K\rangle}\right)
&
=
8 \cdot \operatorname{A}_\text{1-D}\left(\mathbf{T}_{\langle K\rangle}\right)
+
K \cdot \operatorname{A}_\text{1-D}\left(\mathbf{T}_{\langle K\rangle}\right)
\\
&
=
(8+K) \cdot \operatorname{A}_\text{1-D}\left(\mathbf{T}_{\langle K\rangle}\right)
.
\end{split}
\label{general_complex_2d}
\end{equation}
Considering the arithmetic complexity of several DCT approximations
and~\eqref{general_complex_2d},
we assessed the arithmetic complexity of the proposed
pruned approximations.
Results are shown in Table~\ref{tab:complexity}.

\begin{table}%
\centering
\caption{Complexity Assessment}
\begin{tabular}{l| ccc | ccc}
\toprule
\multirow{2}{*}{Method} & \multicolumn{3}{c|}{1-D} & \multicolumn{3}{c}{2-D} \\
& Mult. & Add. & Shift & Mult. & Add. & Shift \\
\midrule
Chen DCT~\cite{Chen1977} & 16 & 26 & 0 & 256 & 416 & 0 \\
SDCT~\cite{haweel2001} & 0 & 24 & 0 & 0 & 384 & 0 \\
BAS 2008~\cite{bas2008} & 0 & 18 & 2 & 0 & 288 & 32 \\
BAS 2009~\cite{bas2009} & 0 & 18 & 0 & 0 & 288 & 0 \\
BAS 2013~\cite{bas2013} & 0 & 24 & 0 & 0 & 284 & 0 \\
RDCT~\cite{cb2011} & 0 & 22 & 0 & 0 & 352 & 0 \\
MRDCT~\cite{bc2012} & 0 & 14 & 0 & 0 & 224 & 0 \\
LODCT~\cite{lengwehasatit2004scalable} & 0 & 24 & 2 & 0 & 384 & 32\\
$\mathbf{T}_4$~\cite{Cintra2014-sigpro} & 0 & 24 & 0 & 0 & 384 & 0 \\
$\mathbf{T}_5$~\cite{Cintra2014-sigpro} & 0 & 24 & 4 & 0 & 384 & 64 \\
$\mathbf{T}_6$~\cite{Cintra2014-sigpro} & 0 & 24 & 6 & 0 & 384 & 96 \\
Proposed $\mathbf{W}_{\langle 4 \rangle}$ & 0 & 18 & 1 & 0 & \textbf{216} & 12 \\
Proposed $\mathbf{M}_{\langle 6 \rangle}$ & 0 & 12 & 0 & 0 & \textbf{168} & 0 \\
\bottomrule
\end{tabular}
\label{tab:complexity}
\end{table}

The proposed $\mathbf{W}_{\langle 4 \rangle}$ has
$43.75~\%$ less operations than the original \mbox{2-D} LODCT.
The proposed $\mathbf{M}_{\langle 6 \rangle}$
attained the lowest additive complexity among the
considered
methods:
12 and 168~additions
for the \mbox{1-D} and \mbox{2-D} cases,
respectively.
Such method presents
$25.0~\%$ less additions than the \mbox{2-D} case of the original MRDCT.
Both pruned methods present lower additive complexity than any of the
considered \mbox{2-D} methods.

\section{Image Compression}

Again
considering the discussed set of images
described in Section~\ref{proposed_pruned_dct},
we
performed a JPEG-like image compression simulation
based on each method
listed in Table~\ref{tab:complexity}~
\cite{haweel2001,bas2008,bas2009}.
Each input image is subdivided into 8$\times$8 blocks,
and each block is submitted to a particular \mbox{2-D}
transformation according to~\eqref{pruned_2d}.
Then,
the resulting coefficients are quantized according
to the standard quantization operation for luminance~\cite[p.~155]{bhaskaran1997}.
The inverse operation is performed to reconstruct the
compressed images.

\begin{figure*}%
\centering
\subfigure[Uncompressed]{\includegraphics[width=0.24\linewidth]{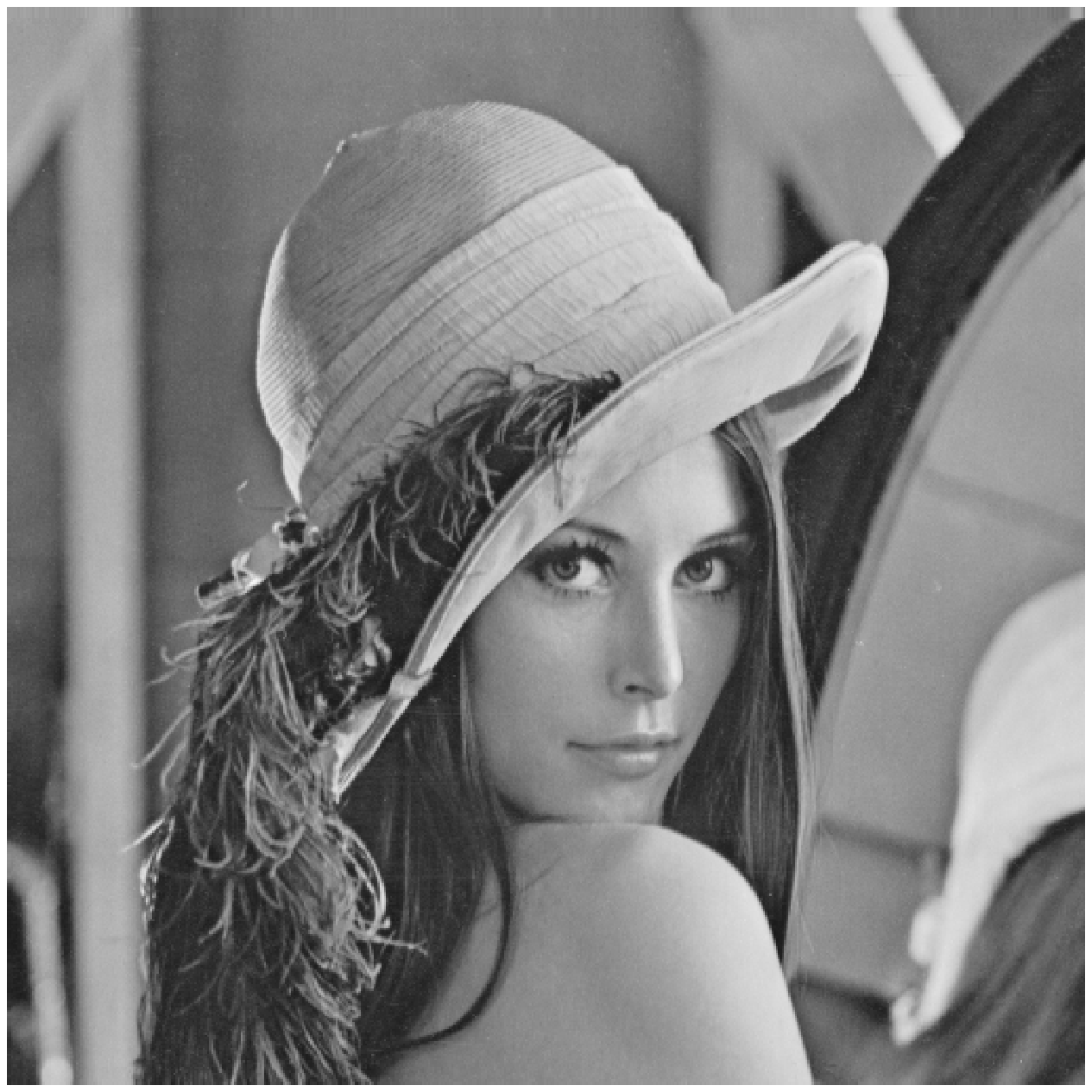}\label{fig:lena_orig}}
\subfigure[Exact DCT (PSNR=35.8279, SSIM= 0.9192)]{\includegraphics[width=0.24\linewidth]{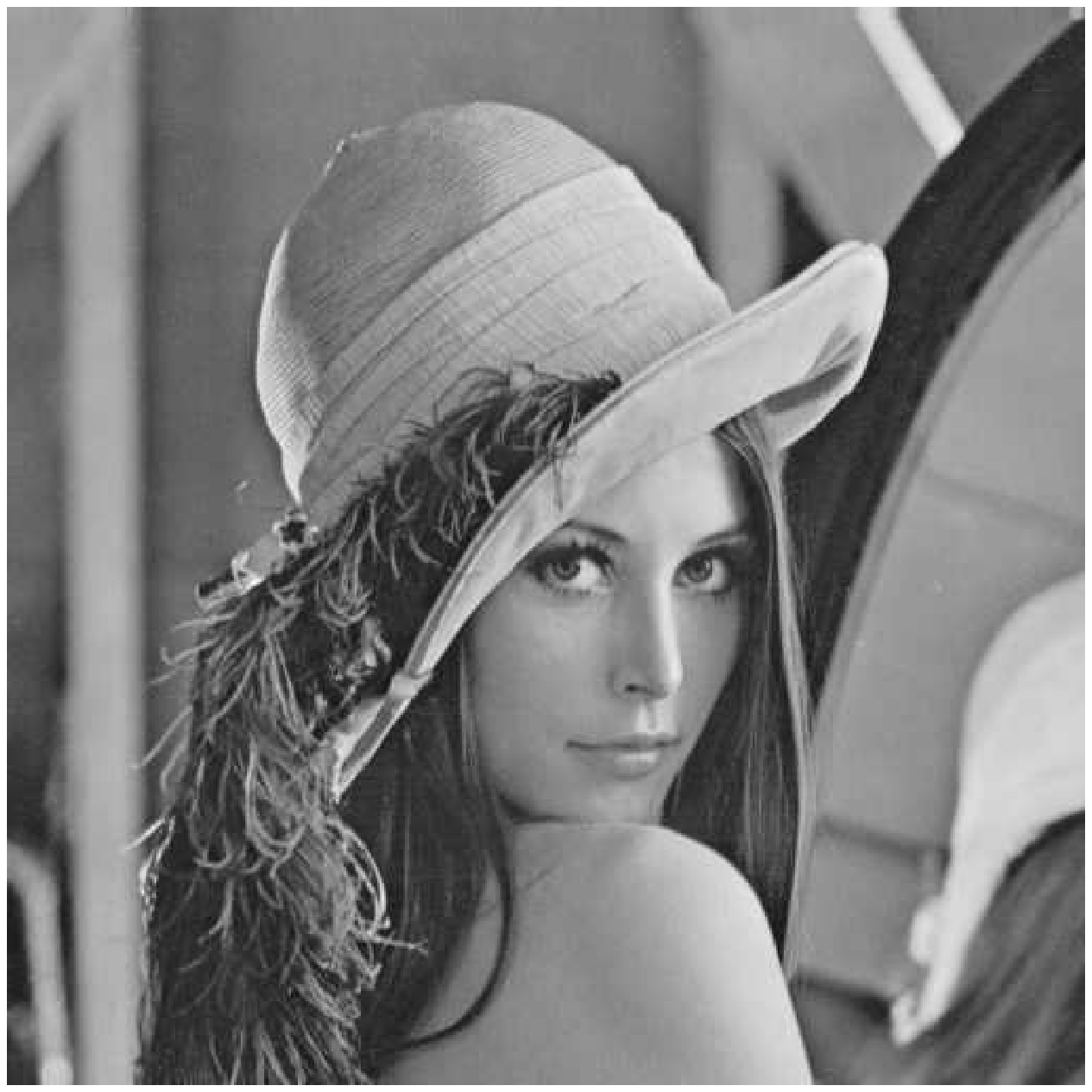}\label{fig:lena_dct}}
\subfigure[Proposed $\mathbf{W}_{\langle 4 \rangle}$ (PSNR=32.1736, SSIM= 0.8855)]{\includegraphics[width=0.24\linewidth]{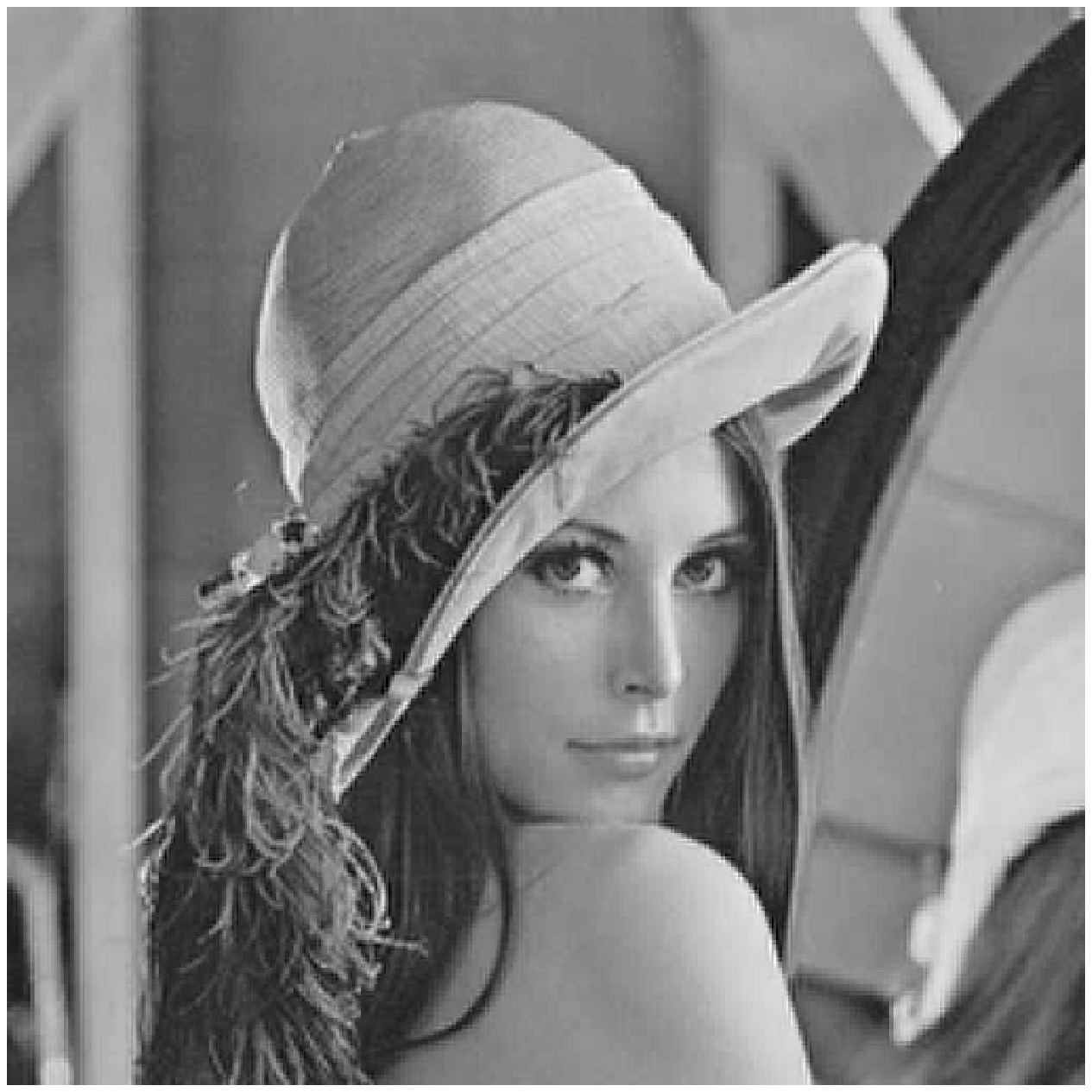}\label{fig:lena_ortegak4}}
\subfigure[Proposed $\mathbf{M}_{\langle 6 \rangle}$ (PSNR=31.6174, SSIM= 0.8647)]{\includegraphics[width=0.24\linewidth]{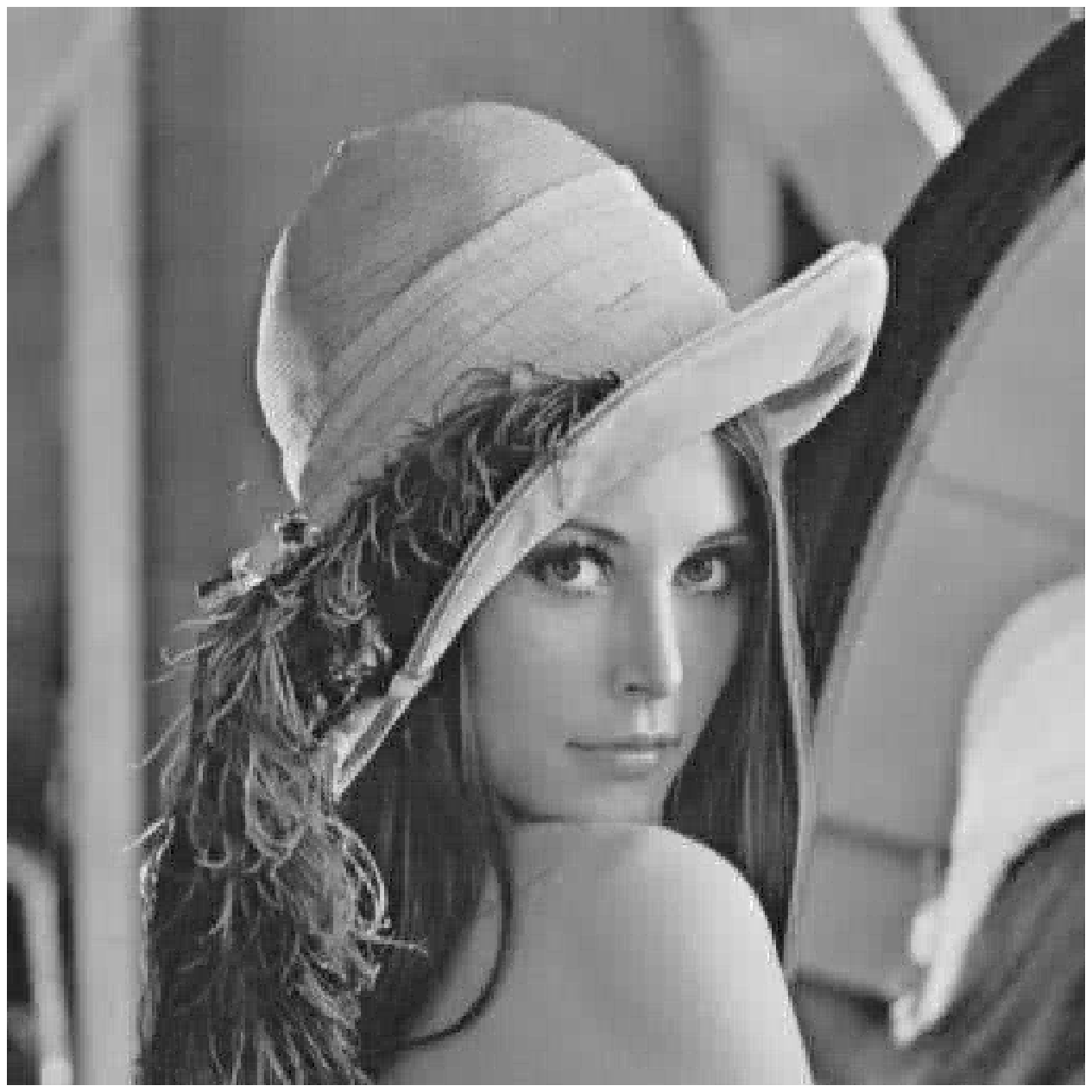}\label{fig:lena_mrdct_k6}} \\
\subfigure[Uncompressed]{\includegraphics[width=0.24\linewidth]{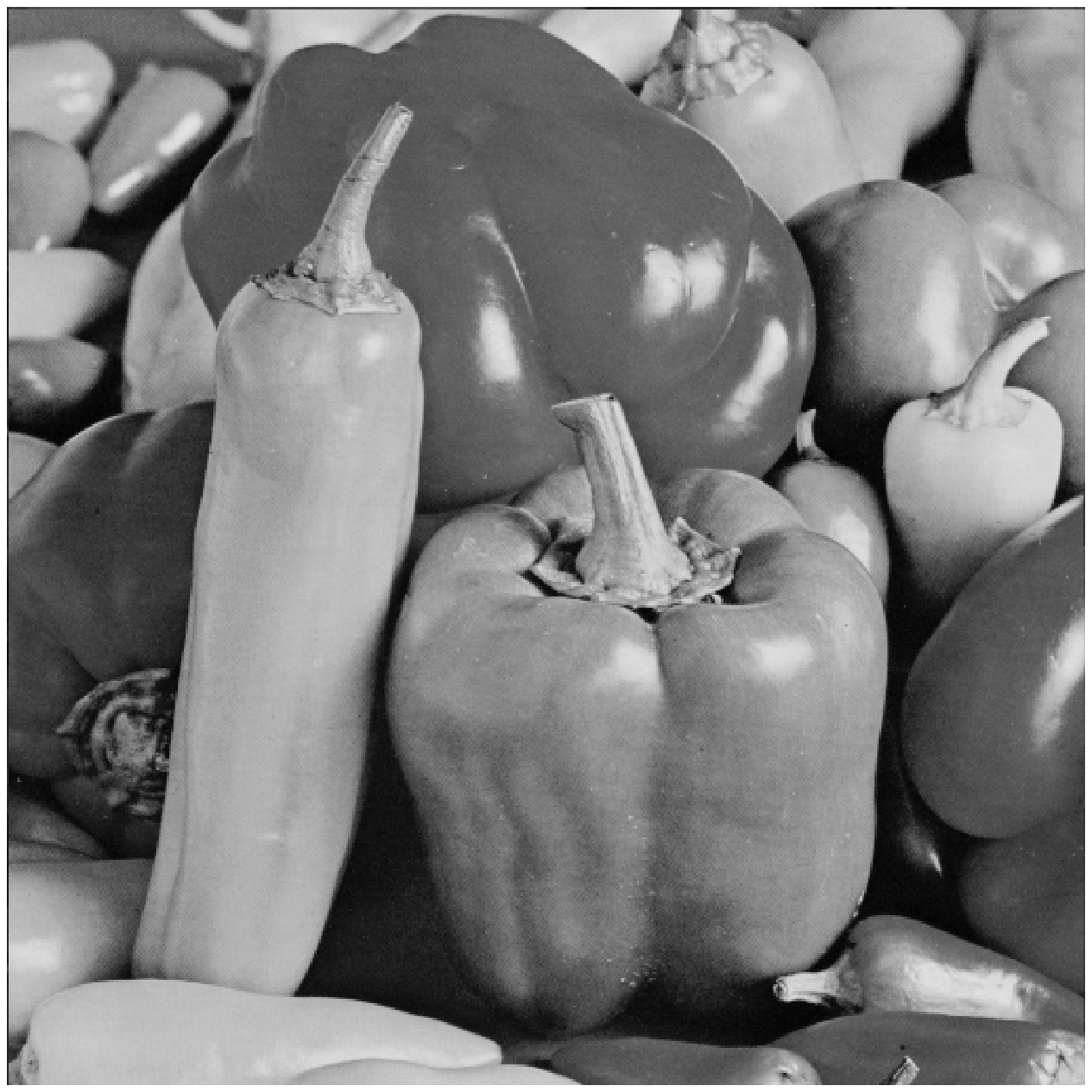}\label{fig:pepper_orig}}
\subfigure[Exact DCT (PSNR=34.7802, SSIM= 0.8814)]{\includegraphics[width=0.24\linewidth]{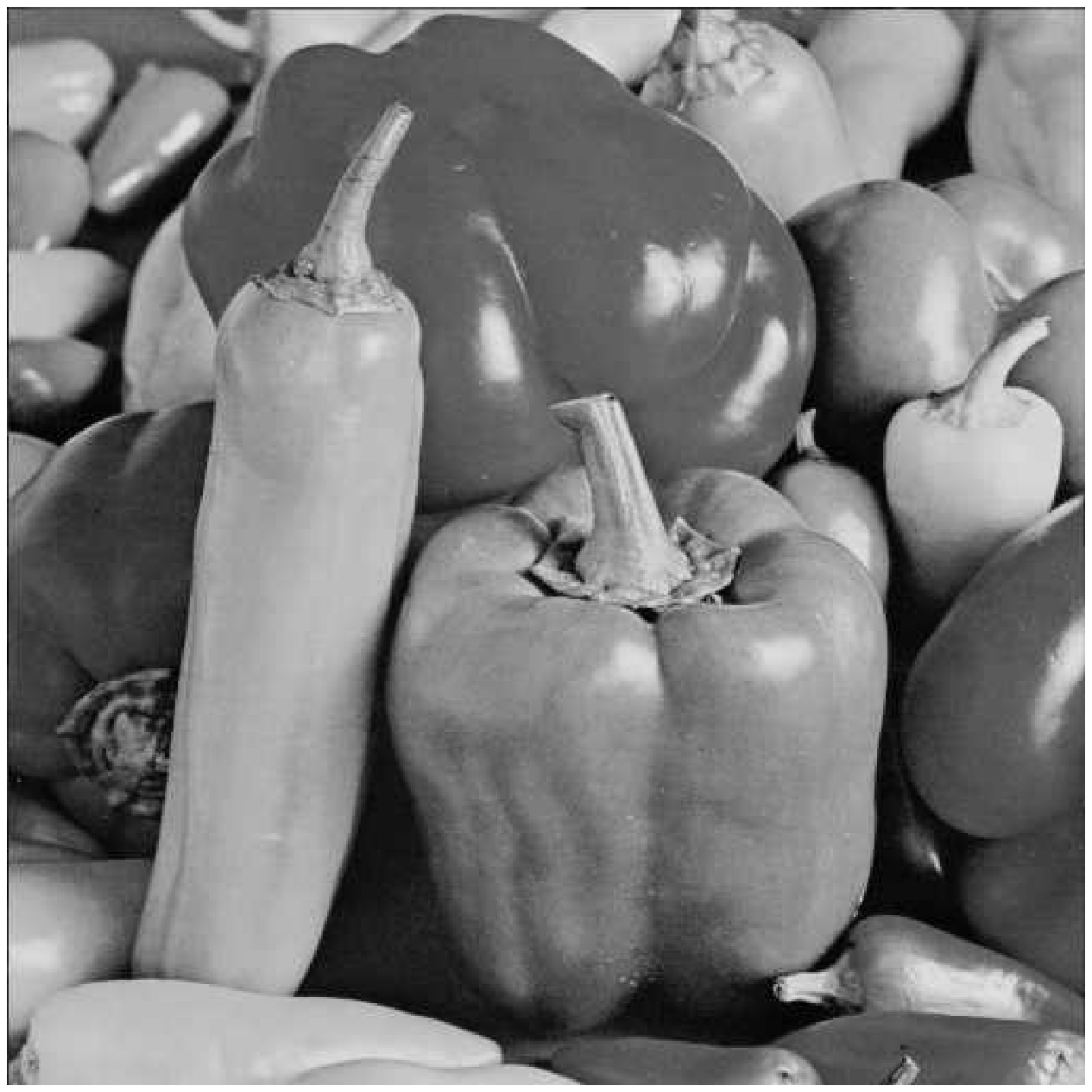}\label{fig:pepper_dct}}
\subfigure[Proposed $\mathbf{W}_{\langle 4 \rangle}$ (PSNR=30.4485, SSIM= 0.8661)]{\includegraphics[width=0.24\linewidth]{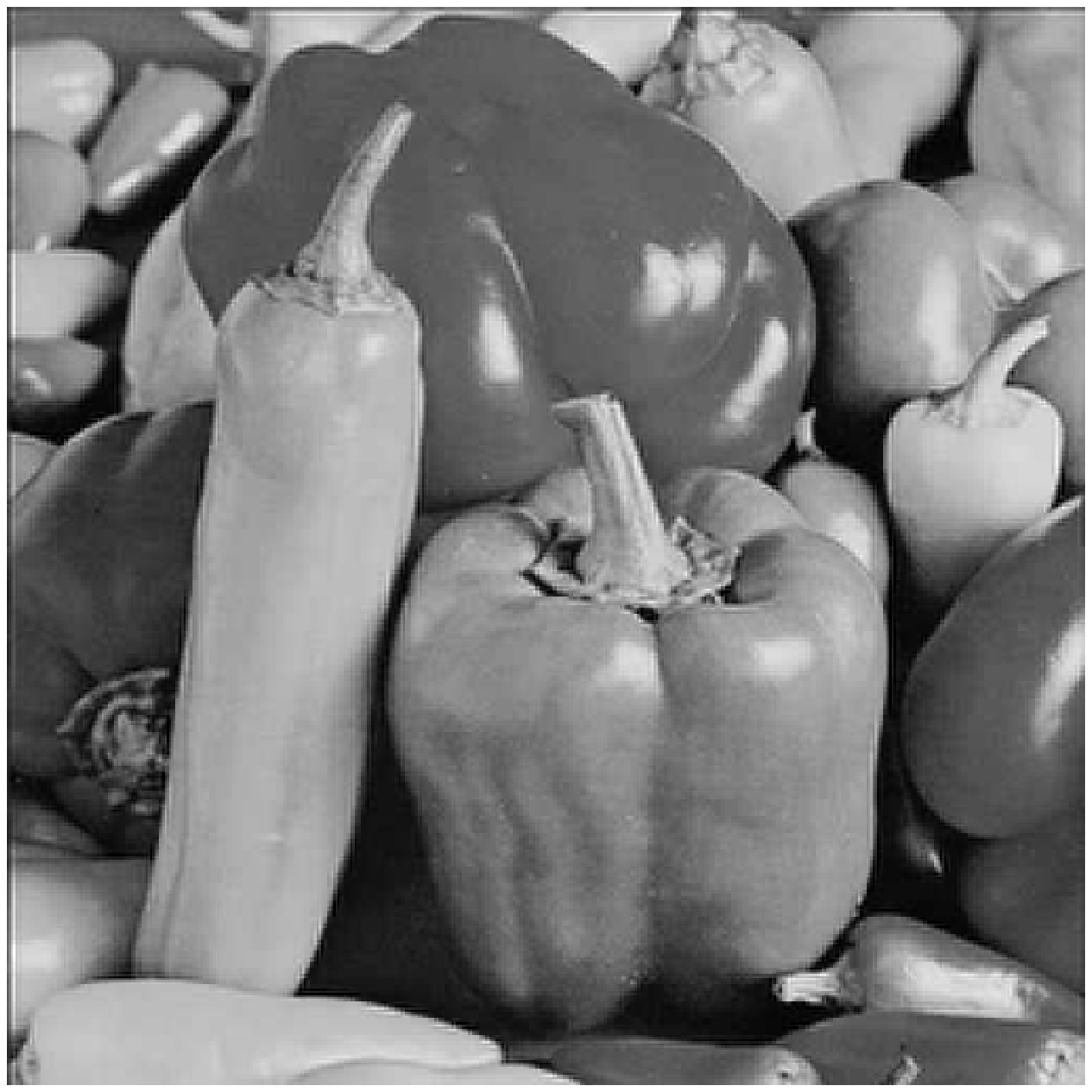}\label{fig:pepper_ortegak4}}
\subfigure[Proposed $\mathbf{M}_{\langle 6 \rangle}$(PSNR=31.5742, SSIM= 0.8380)]{\includegraphics[width=0.24\linewidth]{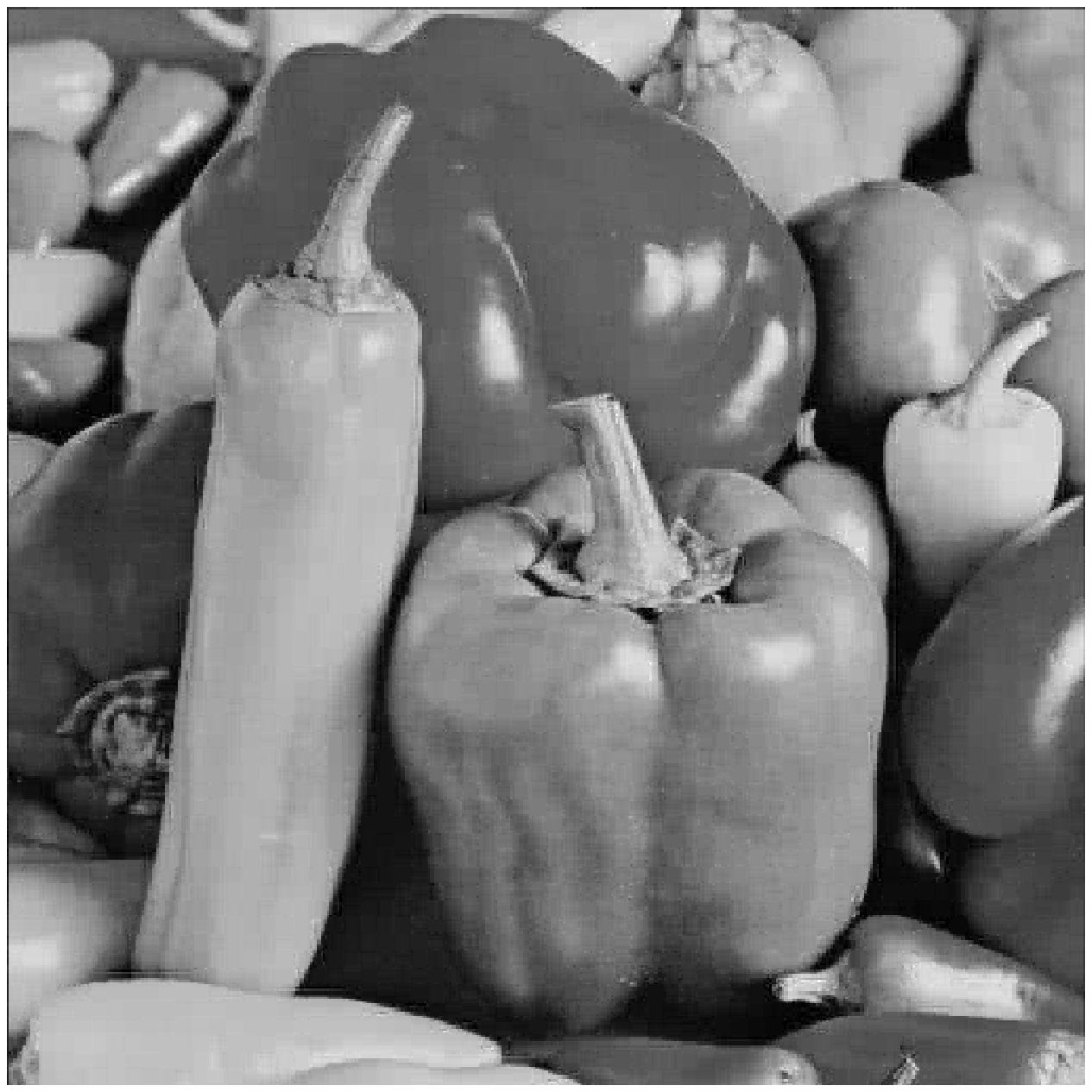}\label{fig:pepper_mrdct_k6}}
\caption{Original and reconstructed Lena and peppers image according to exact DCT and the pruned proposed methods.}
\label{fig:images}
\end{figure*}

Original and reconstructed images are compared quantitatively using the peak signal-to-noise ratio (PSNR)~\cite[p.~9]{bhaskaran1997} and  the structural similarity (SSIM)~\cite{Wang2004} as
figures of merit.
Average measurements were considered.
Table~\ref{tab:performance} shows the results.
\figurename~\ref{fig:images}
shows
a qualitative evaluation
of the proposed methods.
Original uncompressed and compressed Lena images
obtained by means of
the exact DCT and the proposed methods are
depicted.
PSNR and SSIM measurements are included for comparison.
Notice that the such values relate to these particular images,
whereas the measurements presented in Table II consists of the average
values of the considered image set.

\begin{table}%
\centering
\caption{Performance assessment}
\begin{tabular}{l|c|c}
\toprule
Method & PSNR & SSIM \\
\midrule
Exact DCT~\cite{Chen1977} & 33.1276 & 0.9030 \\
SDCT~\cite{haweel2001} & 29.8442 & 0.8356 \\
BAS-2008~\cite{bas2008} & 32.2017 & 0.8851 \\
BAS-2009~\cite{bas2009} & 31.7557 & 0.8763 \\
BAS-2013~\cite{bas2013} & 31.8416 & 0.8815 \\
RDCT~\cite{cb2011} & 31.9553 &  0.8823 \\
MRDCT~\cite{bc2012} & 30.9775 & 0.8552 \\
LODCT.~\cite{lengwehasatit2004scalable} & 32.4364 & 0.8916 \\
$\mathbf{T}_4$~\cite{Cintra2014-sigpro} & 31.9826 & 0.8823 \\
$\mathbf{T}_4$~\cite{Cintra2014-sigpro} & 31.8422 & 0.8796 \\
$\mathbf{T}_5$~\cite{Cintra2014-sigpro} & 32.2892 & 0.8892 \\
Proposed $\mathbf{W}_{\langle 4 \rangle}$ & 29.5282 & 0.8411\\
Proposed $\mathbf{M}_{\langle 6 \rangle}$ & 29.5810 & 0.8349\\
\bottomrule
\end{tabular}
\label{tab:performance}
\end{table}

\section{VLSI Architectures}

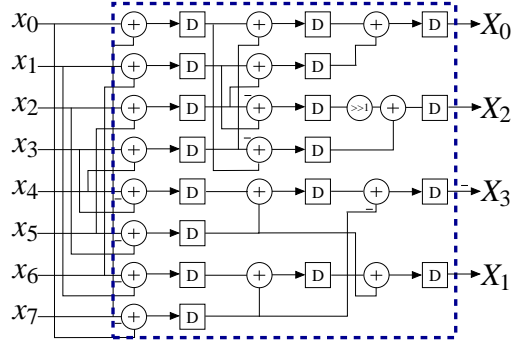
\begin{figure}
\centering
\scalebox{0.70}{\input{vitorpaperfig2.pstex_t}}
\caption{\mbox{1-D} architecture of the proposed 8-point pruned
LODCT.}
\label{prunedDCT}
\end{figure}

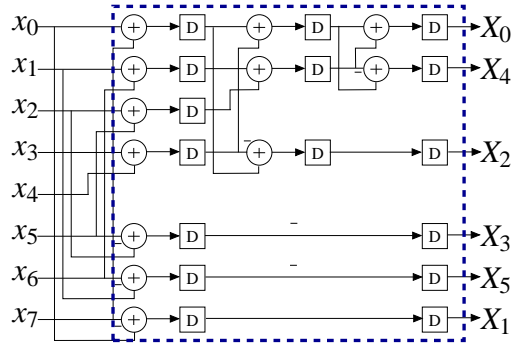
\begin{figure}
\centering
\scalebox{0.70}{\input{vitorpaperfig1.pstex_t}}
\caption{\mbox{1-D} architecture of the proposed 8-point pruned MRDCT.}
\label{MRDCT}
\end{figure}

In this section,
hardware architectures
for the proposed pruned LODCT
and
pruned MRDCT are detailed.
The \mbox{1-D} version of each transformation
were initially modeled and tested in Matlab Simulink.
\figurename~\ref{prunedDCT} and~\ref{MRDCT} depict
the resulting architectures for the pruned LODCT and pruned MRDCT,
respectively.

\subsection{FPGA Implementations}

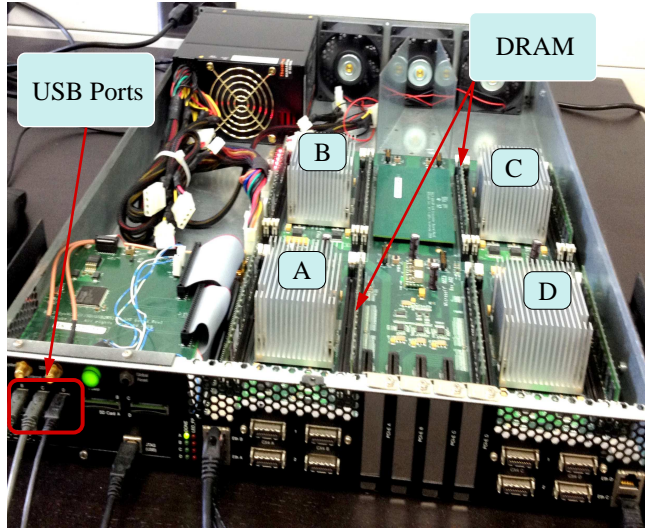
\begin{figure}
\centering
\scalebox{0.55}{\input{vitorpaperfig3.pstex_t}}
\caption{BEE3 device used to implement the architectures.
Labels A, B, C, and D indicate
the four Xilinx Virtex-5 XC5VSX95T-2FF1136 FPGA devices.}
\label{BEE3}
\end{figure}

The above discussed architectures were physically realized
on Berkeley Emulation Engine~(BEE3)~\cite{bee3},
a multi-FPGA based rapid prototyping system and was
tested using on-chip hardware-in-the-loop co-simulation.
The BEE3 system consists of a 2U chassis with a tightly-couple four FPGA system,
widely employed in academia and industry.
The main printed circuit Board~(PCB) and
a control \& I/O PCB supports four Xilinx Virtex 5 FPGAs,
up to 16~DDR2 DIMMs,
eight 10GBase-CX4 Interfaces,
four PCI-Express slots,
four USB ports,
four 1GbE RJ45 Ports,
and
one Xilinx USB-JTAG Interface,
as illustrated in
\figurename~\ref{BEE3}.
Such device was employed
to physically realize the above architectures
with fine-grain pipelining for increased throughput.
FPGA realizations were tested with
10{,}000 random 8-point input test vectors.
Test vectors were generated from
within the MATLAB environment
and
routed to the BEE3 device through the USB ports and then
measured data from the BEE3 device was routed back
to MATLAB memory space.

Evaluation of hardware complexity and real time performance
considered the following metrics:
the number of used configurable logic blocks (CLB),
flip-flop (FF) count,
critical path delay ($T_\text{cpd}$),
and
the maximum operating frequency ($F_{\text{max}}$) in~MHz.
The \texttt{xflow.results} report file
was accessed to obtain the above results.
Dynamic ($D_p$)
and
static power ($Q_p$) consumptions
were estimated using the Xilinx XPower Analyzer
for the Xilinx Virtex-5 XC5VSX95T-2FF1136 device.
Results are shown in Table~\ref{FPGAresults}.

\begin{table}
\centering
\caption{Hardware resource consumption using Xilinx Virtex-5 XC5VSX95T-2FF1136 device}
\begin{tabular}{l c|c}
\toprule
 & \multicolumn{2}{c}{Method} \\
\cmidrule{2-3}
Hardware metric & Pruned LODCT & Pruned MRDCT \\
\midrule
CLB & 298 & 232 \\
FF & 1054 & 895 \\
$T_\text{cpd}$ ($\mathrm{ns}$) & 3.578 & 3.588 \\
$F_{\text{max}}$ ($\mathrm{MHz}$) & 279.48 & 278.70 \\
$D_p$ ($\mathrm{mW/MHz}$) & 3.141 & 3.620 \\
$Q_p$ ($\mathrm{W}$) & 1.50 & 1.50 \\
\bottomrule
\end{tabular}
\label{FPGAresults}
\end{table}

\section{Conclusion}

In this paper,
we introduced two pruning-based DCT approximations
with very low arithmetic complexity.
The resulting transformations require
lower additive complexity
than state-of-the-art methods.
An image compression simulation were performed.
Quantitative and qualitative assessments
according to well-established figures of merits
indicate
the adequateness of the proposed methods.
VLSI hardware realizations
were proposed
demonstrating the practicability of the proposed approximations.

\section*{Acknowledgment}

The authors would like to thank CNPq, FACEPE, FAPERGS,
and The University of Akron.

{\small
\bibliographystyle{IEEEtran}
\bibliography{ref}
}

\end{document}

%% file: new_ortegak4.pstex_t
\begin{picture}(0,0)%
\includegraphics{new_ortegak4.pstex}%
\end{picture}%
\setlength{\unitlength}{3937sp}%
\begingroup\makeatletter\ifx\SetFigFont\undefined%
\gdef\SetFigFont#1#2#3#4#5{%
  \reset@font\fontsize{#1}{#2pt}%
  \fontfamily{#3}\fontseries{#4}\fontshape{#5}%
  \selectfont}%
\fi\endgroup%
\begin{picture}(3234,2357)(710,-2058)
\put(3929,164){\makebox(0,0)[lb]{\smash{{\SetFigFont{8}{9.6}{\rmdefault}{\mddefault}{\updefault}{\color[rgb]{0,0,0}$X_0$}%
}}}}
\put(3929,-454){\makebox(0,0)[lb]{\smash{{\SetFigFont{8}{9.6}{\rmdefault}{\mddefault}{\updefault}{\color[rgb]{0,0,0}$X_2$}%
}}}}
\put(3929,-145){\makebox(0,0)[lb]{\smash{{\SetFigFont{8}{9.6}{\rmdefault}{\mddefault}{\updefault}{\color[rgb]{0,0,0}$X_1$}%
}}}}
\put(725,-145){\makebox(0,0)[lb]{\smash{{\SetFigFont{8}{9.6}{\rmdefault}{\mddefault}{\updefault}{\color[rgb]{0,0,0}$x_1$}%
}}}}
\put(725,-454){\makebox(0,0)[lb]{\smash{{\SetFigFont{8}{9.6}{\rmdefault}{\mddefault}{\updefault}{\color[rgb]{0,0,0}$x_2$}%
}}}}
\put(725,-763){\makebox(0,0)[lb]{\smash{{\SetFigFont{8}{9.6}{\rmdefault}{\mddefault}{\updefault}{\color[rgb]{0,0,0}$x_3$}%
}}}}
\put(725,-1072){\makebox(0,0)[lb]{\smash{{\SetFigFont{8}{9.6}{\rmdefault}{\mddefault}{\updefault}{\color[rgb]{0,0,0}$x_4$}%
}}}}
\put(725,-1380){\makebox(0,0)[lb]{\smash{{\SetFigFont{8}{9.6}{\rmdefault}{\mddefault}{\updefault}{\color[rgb]{0,0,0}$x_5$}%
}}}}
\put(725,-1689){\makebox(0,0)[lb]{\smash{{\SetFigFont{8}{9.6}{\rmdefault}{\mddefault}{\updefault}{\color[rgb]{0,0,0}$x_6$}%
}}}}
\put(725,-1998){\makebox(0,0)[lb]{\smash{{\SetFigFont{8}{9.6}{\rmdefault}{\mddefault}{\updefault}{\color[rgb]{0,0,0}$x_7$}%
}}}}
\put(725,164){\makebox(0,0)[lb]{\smash{{\SetFigFont{8}{9.6}{\rmdefault}{\mddefault}{\updefault}{\color[rgb]{0,0,0}$x_0$}%
}}}}
\put(2791,-331){\makebox(0,0)[lb]{\smash{{\SetFigFont{8}{9.6}{\rmdefault}{\mddefault}{\updefault}{\color[rgb]{0,0,0}$1/2$}%
}}}}
\put(3929,-763){\makebox(0,0)[lb]{\smash{{\SetFigFont{8}{9.6}{\rmdefault}{\mddefault}{\updefault}{\color[rgb]{0,0,0}$X_3$}%
}}}}
\end{picture}%

%% file: mrdct_k6.pstex_t
\begin{picture}(0,0)%
\includegraphics{mrdct_k6.pstex}%
\end{picture}%
\setlength{\unitlength}{4144sp}%
\begingroup\makeatletter\ifx\SetFigFont\undefined%
\gdef\SetFigFont#1#2#3#4#5{%
  \reset@font\fontsize{#1}{#2pt}%
  \fontfamily{#3}\fontseries{#4}\fontshape{#5}%
  \selectfont}%
\fi\endgroup%
\begin{picture}(2460,1770)(301,-1021)
\put(316,614){\makebox(0,0)[lb]{\smash{{\SetFigFont{8}{9.6}{\rmdefault}{\mddefault}{\updefault}{\color[rgb]{0,0,0}$x_0$}%
}}}}
\put(316,164){\makebox(0,0)[lb]{\smash{{\SetFigFont{8}{9.6}{\rmdefault}{\mddefault}{\updefault}{\color[rgb]{0,0,0}$x_2$}%
}}}}
\put(316,-61){\makebox(0,0)[lb]{\smash{{\SetFigFont{8}{9.6}{\rmdefault}{\mddefault}{\updefault}{\color[rgb]{0,0,0}$x_3$}%
}}}}
\put(316,-286){\makebox(0,0)[lb]{\smash{{\SetFigFont{8}{9.6}{\rmdefault}{\mddefault}{\updefault}{\color[rgb]{0,0,0}$x_4$}%
}}}}
\put(316,-511){\makebox(0,0)[lb]{\smash{{\SetFigFont{8}{9.6}{\rmdefault}{\mddefault}{\updefault}{\color[rgb]{0,0,0}$x_5$}%
}}}}
\put(316,-736){\makebox(0,0)[lb]{\smash{{\SetFigFont{8}{9.6}{\rmdefault}{\mddefault}{\updefault}{\color[rgb]{0,0,0}$x_6$}%
}}}}
\put(316,-961){\makebox(0,0)[lb]{\smash{{\SetFigFont{8}{9.6}{\rmdefault}{\mddefault}{\updefault}{\color[rgb]{0,0,0}$x_7$}%
}}}}
\put(316,389){\makebox(0,0)[lb]{\smash{{\SetFigFont{8}{9.6}{\rmdefault}{\mddefault}{\updefault}{\color[rgb]{0,0,0}$x_1$}%
}}}}
\put(2746,-61){\makebox(0,0)[lb]{\smash{{\SetFigFont{8}{9.6}{\rmdefault}{\mddefault}{\updefault}{\color[rgb]{0,0,0}$X_2$}%
}}}}
\put(2746,-511){\makebox(0,0)[lb]{\smash{{\SetFigFont{8}{9.6}{\rmdefault}{\mddefault}{\updefault}{\color[rgb]{0,0,0}$X_3$}%
}}}}
\put(2746,-736){\makebox(0,0)[lb]{\smash{{\SetFigFont{8}{9.6}{\rmdefault}{\mddefault}{\updefault}{\color[rgb]{0,0,0}$X_5$}%
}}}}
\put(2746,-961){\makebox(0,0)[lb]{\smash{{\SetFigFont{8}{9.6}{\rmdefault}{\mddefault}{\updefault}{\color[rgb]{0,0,0}$X_1$}%
}}}}
\put(2746,389){\makebox(0,0)[lb]{\smash{{\SetFigFont{8}{9.6}{\rmdefault}{\mddefault}{\updefault}{\color[rgb]{0,0,0}$X_4$}%
}}}}
\put(2746,614){\makebox(0,0)[lb]{\smash{{\SetFigFont{8}{9.6}{\rmdefault}{\mddefault}{\updefault}{\color[rgb]{0,0,0}$X_0$}%
}}}}
\end{picture}%

%% file: vitorpaperfig2.pstex_t
\begin{picture}(0,0)%
\includegraphics{vitorpaperfig2.pstex}%
\end{picture}%
\setlength{\unitlength}{3947sp}%
\begingroup\makeatletter\ifx\SetFigFont\undefined%
\gdef\SetFigFont#1#2#3#4#5{%
  \reset@font\fontsize{#1}{#2pt}%
  \fontfamily{#3}\fontseries{#4}\fontshape{#5}%
  \selectfont}%
\fi\endgroup%
\begin{picture}(4236,3066)(136,-2269)
\put(151,179){\makebox(0,0)[lb]{\smash{{\SetFigFont{17}{20.4}{\rmdefault}{\mddefault}{\updefault}{\color[rgb]{0,0,0}$x_{1}$}%
}}}}
\put(151,-196){\makebox(0,0)[lb]{\smash{{\SetFigFont{17}{20.4}{\rmdefault}{\mddefault}{\updefault}{\color[rgb]{0,0,0}$x_{2}$}%
}}}}
\put(151,-571){\makebox(0,0)[lb]{\smash{{\SetFigFont{17}{20.4}{\rmdefault}{\mddefault}{\updefault}{\color[rgb]{0,0,0}$x_{3}$}%
}}}}
\put(151,-946){\makebox(0,0)[lb]{\smash{{\SetFigFont{17}{20.4}{\rmdefault}{\mddefault}{\updefault}{\color[rgb]{0,0,0}$x_{4}$}%
}}}}
\put(151,-1321){\makebox(0,0)[lb]{\smash{{\SetFigFont{17}{20.4}{\rmdefault}{\mddefault}{\updefault}{\color[rgb]{0,0,0}$x_{5}$}%
}}}}
\put(151,-1696){\makebox(0,0)[lb]{\smash{{\SetFigFont{17}{20.4}{\rmdefault}{\mddefault}{\updefault}{\color[rgb]{0,0,0}$x_{6}$}%
}}}}
\put(151,-2071){\makebox(0,0)[lb]{\smash{{\SetFigFont{17}{20.4}{\rmdefault}{\mddefault}{\updefault}{\color[rgb]{0,0,0}$x_{7}$}%
}}}}
\put(151,554){\makebox(0,0)[lb]{\smash{{\SetFigFont{17}{20.4}{\rmdefault}{\mddefault}{\updefault}{\color[rgb]{0,0,0}$x_{0}$}%
}}}}
\put(4355,489){\makebox(0,0)[lb]{\smash{{\SetFigFont{17}{20.4}{\rmdefault}{\mddefault}{\updefault}{\color[rgb]{0,0,0}$X_{0}$}%
}}}}
\put(4355,-261){\makebox(0,0)[lb]{\smash{{\SetFigFont{17}{20.4}{\rmdefault}{\mddefault}{\updefault}{\color[rgb]{0,0,0}$X_{2}$}%
}}}}
\put(4356,-1012){\makebox(0,0)[lb]{\smash{{\SetFigFont{17}{20.4}{\rmdefault}{\mddefault}{\updefault}{\color[rgb]{0,0,0}$X_{3}$}%
}}}}
\put(4356,-1762){\makebox(0,0)[lb]{\smash{{\SetFigFont{17}{20.4}{\rmdefault}{\mddefault}{\updefault}{\color[rgb]{0,0,0}$X_{1}$}%
}}}}
\end{picture}%

%% file: vitorpaperfig1.pstex_t
\begin{picture}(0,0)%
\includegraphics{vitorpaperfig1.pstex}%
\end{picture}%
\setlength{\unitlength}{3947sp}%
\begingroup\makeatletter\ifx\SetFigFont\undefined%
\gdef\SetFigFont#1#2#3#4#5{%
  \reset@font\fontsize{#1}{#2pt}%
  \fontfamily{#3}\fontseries{#4}\fontshape{#5}%
  \selectfont}%
\fi\endgroup%
\begin{picture}(4235,3066)(136,-3169)
\put(151,-346){\makebox(0,0)[lb]{\smash{{\SetFigFont{17}{20.4}{\rmdefault}{\mddefault}{\updefault}{\color[rgb]{0,0,0}$x_{0}$}%
}}}}
\put(4355,-786){\makebox(0,0)[lb]{\smash{{\SetFigFont{17}{20.4}{\rmdefault}{\mddefault}{\updefault}{\color[rgb]{0,0,0}$X_{4}$}%
}}}}
\put(4355,-2286){\makebox(0,0)[lb]{\smash{{\SetFigFont{17}{20.4}{\rmdefault}{\mddefault}{\updefault}{\color[rgb]{0,0,0}$X_{3}$}%
}}}}
\put(4355,-2661){\makebox(0,0)[lb]{\smash{{\SetFigFont{17}{20.4}{\rmdefault}{\mddefault}{\updefault}{\color[rgb]{0,0,0}$X_{5}$}%
}}}}
\put(4355,-3036){\makebox(0,0)[lb]{\smash{{\SetFigFont{17}{20.4}{\rmdefault}{\mddefault}{\updefault}{\color[rgb]{0,0,0}$X_{1}$}%
}}}}
\put(4355,-1536){\makebox(0,0)[lb]{\smash{{\SetFigFont{17}{20.4}{\rmdefault}{\mddefault}{\updefault}{\color[rgb]{0,0,0}$X_{2}$}%
}}}}
\put(4355,-411){\makebox(0,0)[lb]{\smash{{\SetFigFont{17}{20.4}{\rmdefault}{\mddefault}{\updefault}{\color[rgb]{0,0,0}$X_{0}$}%
}}}}
\put(151,-721){\makebox(0,0)[lb]{\smash{{\SetFigFont{17}{20.4}{\rmdefault}{\mddefault}{\updefault}{\color[rgb]{0,0,0}$x_{1}$}%
}}}}
\put(151,-1096){\makebox(0,0)[lb]{\smash{{\SetFigFont{17}{20.4}{\rmdefault}{\mddefault}{\updefault}{\color[rgb]{0,0,0}$x_{2}$}%
}}}}
\put(151,-1471){\makebox(0,0)[lb]{\smash{{\SetFigFont{17}{20.4}{\rmdefault}{\mddefault}{\updefault}{\color[rgb]{0,0,0}$x_{3}$}%
}}}}
\put(151,-1846){\makebox(0,0)[lb]{\smash{{\SetFigFont{17}{20.4}{\rmdefault}{\mddefault}{\updefault}{\color[rgb]{0,0,0}$x_{4}$}%
}}}}
\put(151,-2221){\makebox(0,0)[lb]{\smash{{\SetFigFont{17}{20.4}{\rmdefault}{\mddefault}{\updefault}{\color[rgb]{0,0,0}$x_{5}$}%
}}}}
\put(151,-2596){\makebox(0,0)[lb]{\smash{{\SetFigFont{17}{20.4}{\rmdefault}{\mddefault}{\updefault}{\color[rgb]{0,0,0}$x_{6}$}%
}}}}
\put(151,-2971){\makebox(0,0)[lb]{\smash{{\SetFigFont{17}{20.4}{\rmdefault}{\mddefault}{\updefault}{\color[rgb]{0,0,0}$x_{7}$}%
}}}}
\end{picture}%

%% file: vitorpaperfig3.pstex_t
\begin{picture}(0,0)%
\includegraphics{vitorpaperfig3.pstex}%
\end{picture}%
\setlength{\unitlength}{3947sp}%
\begingroup\makeatletter\ifx\SetFigFont\undefined%
\gdef\SetFigFont#1#2#3#4#5{%
  \reset@font\fontsize{#1}{#2pt}%
  \fontfamily{#3}\fontseries{#4}\fontshape{#5}%
  \selectfont}%
\fi\endgroup%
\begin{picture}(7299,6000)(-23,-5161)
\end{picture}%